\documentclass[prd,aps,graphics]{revtex4}
\usepackage{epsfig}

\newcommand{\dd}{{\rm d}}
\newcommand{\bk}{{\bf k}}
\newcommand{\bx}{{\bf x}}

\newcommand{\chib}{\overline{\chi}}
\newcommand{\vchi}{{v}}
\newcommand{\mg}{\langle}
\newcommand{\md}{\rangle}
\def\vk{{\bf k}}
\def\vx{{\bf x}}
\def\d{{\rm d}}
\def\dd{{\rm d}}
\def\ii{{\rm i}}

\begin{document}
\title{High order correlation functions for self interacting
scalar field in de Sitter space}

\author{Francis Bernardeau}
\affiliation{Service de Physique Th{\'e}orique,
         CEA/DSM/SPhT, Unit{\'e} de recherche associ{\'e}e au CNRS, CEA/Saclay
         91191 Gif-sur-Yvette c{\'e}dex}
\author{Tristan Brunier}
\affiliation{Service de Physique Th{\'e}orique,
         CEA/DSM/SPhT, Unit{\'e} de recherche associ{\'e}e au CNRS, CEA/Saclay
         91191 Gif-sur-Yvette c{\'e}dex}
\author{Jean--Philippe Uzan}
\affiliation{Laboratoire de Physique Th{\'e}orique, CNRS--UMR 8627,
         B{\^a}t. 210, Universit{\'e} Paris XI, F--91405 Orsay Cedex,
         France,\\ and \\
         Institut d'Astrophysique de Paris, GReCO,
        CNRS-FRE 2435, 98 bis, Bd Arago, 75014 Paris, France.}

\vskip 0.15cm

\date{Notes du 30 Octobre 2003}

\begin{abstract}

We present the expressions of the three- and four-point correlation
functions of a self interacting light scalar field in a de Sitter
spacetime at tree order respectively for a cubic and a quartic
potential. Exact expressions are derived and their limiting behaviour
on super-horizon scales are presented. Their essential features are
shown to be similar to those obtained in a classical approach.

\end{abstract}
\pacs{{ \bf PACS numbers:} } \vskip2pc

\maketitle
\section{Introduction}

The computation of high order correlation functions of scalar fields
in a de Sitter spacetime can be of interest for investigations of the
physics of the early universe. There are indeed a growing number of
indications that in its early phase the universe underwent an
inflationary period~\cite{WMAP} that can be accurately described by a
de Sitter phase during which the energy density of the universe is
thought to be dominated by the self energy of a scalar
field~\cite{guth81}. Its fluctuations, or the fluctuations of any
other light scalar fields, are thought to be the progenitors of the
large-scale structure of the universe~\cite{inflation}. The
statistical properties of the induced metric fluctuations then depend
on the potential landscape in which the fields evolve. It is unlikely
that the fluctuations produced along the inflaton direction can be
significantly non-Gaussian as stressed in recent
works~\cite{maldacena,Creminelli}. It is possible however that if
there exist other self interacting light scalar fields during that
period primordial non Gaussian metric fluctuations can be
generated~\cite{bartolo2,bartolo1,BU1,BU2}. The details of the
effects induced by such self-interacting fields is based on the
statistical properties of the metric fluctuations that quantum
generated field fluctuations can induced.

The purpose of this paper is to investigate the foundation of
these calculations by calculating the correlation properties of a
test scalar field in a de Sitter background.

The content of the paper is the following. In the second part we
present the basis of such computations. In the third we explore the
expression of the leading order term of high order correlators for
different self-interacting potentials. In particular we give explicit
results for the superhorizon limit that corresponds to modes that can
be observed today. Finally we give insights on the physical
interpretations of those results.

\section{Correlation function computations}

\subsection{Free field behavior}

For the purpose of our calculations we assume that the
inflationary phase can be described by a de Sitter background
epoch. This description is only approximate since during the
inflationary period the Hubble constant slowly varies with time
but it is a framework in which all calculations can be pursued
analytically. We have checked that to a large extent the results
would not be affected if the background is
changed~\footnote{Calculations in other background are difficult in
general but we have checked that in case of a power law inflation,
$a(t)\propto t^{\nu}$, we recover the same behaviors in the
superhorizon limit if $\nu$ is large enough.}.

A de Sitter spacetime~\cite{hawking} in flat spatial section
slicing is described by the metric,
\begin{equation}\label{eq:m}
\dd s^2=\frac{1}{(H\eta)^2}\left(-\dd\eta^2+\delta_{ij}\dd x^i\dd
x^j\right)
\end{equation}
which is conformal to half of the Minkowski spacetime. The
conformal time is related to the cosmological time by
\begin{equation}
\eta=-\frac{1}{H}\hbox{e}^{-Ht}
\end{equation}
and runs from $-\infty$ to 0, the limit $\eta\rightarrow0^-$
representing the ``infinite future''.

For a minimally coupled free quantum field of mass $m$, the solution
can be decomposed in plane waves as
\begin{equation}\label{eq:quantdec}
 \widehat \vchi_0(\bx,\eta)=\int\dd^3\bk
 \left[\vchi_0(k,\eta)\widehat b_\bk\hbox{e}^{i\bk\cdot\bx}
 +\vchi^*_0(k,\eta)\widehat b_\bk^\dag\hbox{e}^{-i\bk\cdot\bx}\right]
\end{equation}
where we have introduced $\widehat \vchi\equiv a\,\widehat\chi$, a
hat referring to a quantum operator. In this Heisenberg picture,
the field has become a time-dependent operator expanded in terms
of time-independent creation and annihilation operators satisfying
the usual commutation relations $[\widehat b_\bk,\widehat
b_{\bk'}^\dag]=\delta_{\rm Dirac}(\bk-\bk')$. We can then define
the free vacuum state by the requirement
\begin{equation}
 \widehat b_{\bf k}\left|0\right>=0\qquad \hbox{for all}\ {\bf k}.
\end{equation}
As it is standard while working in curved spacetime~\cite{birrel}, the
definition of the vacuum state suffers from some arbitrariness
since it depends on the choice of the set of modes
$\chi_0(k,\eta)$. They satisfy the evolution equation
\begin{equation}\label{eq:v}
  \vchi_0''+\left(k^2-\frac{2}{\eta^2}-\frac{m^2/H^2}{\eta^2}\right)\vchi_0=0,
\end{equation}
the general solution of which is given by
$\sqrt{\pi\eta/4}\left[c_1H^{(1)}_\nu(k\eta)+c_2H^{(2)}_\nu(k\eta)\right]$
with $|c_2|^2-|c_1|^2=1$, where $H^{(1)}_\nu$ and $H^{(2)}_\nu$
are the Hankel functions of first and second kind and with
$\nu^2=9/4-m^2/H^2$. Among this family of solutions, it is natural
to choose the one enjoying the de Sitter symmetry and the same
short distance behavior than in Minkowski spacetime. This leads to
\begin{equation}\label{eq:sol}
\vchi_0(k,\eta)= \frac{1}{2}\sqrt{\pi\eta}H^{(2)}_\nu(k\eta).
\end{equation}
This uniquely defines a de Sitter invariant vacuum state referred
to as the Bunch-Davies state vacuum~\cite{birrel}. In the massless
limit, the solution (\ref{eq:v}) reduces to
\begin{equation}
\vchi_0(k,\eta)=\left(1-\frac{\ii}{k\eta}\right)\frac{\hbox{e}^{-\ii
k\eta}}{\sqrt{2k}}.
\end{equation}
This result gives the expression of the equal-time two-point
correlator of the Fourier modes,
\begin{eqnarray}
  \mg \chi(\vk_1)\chi(\vk_2)\md &=& \delta_{\rm Dirac}(\vk_1+\vk_2)
  \,P_2(k_1)\\
  P_2(k_1)&=&\frac{H^2\eta^2}{2k_1}\left(1+
  \frac{1}{k_1^2\eta^2}\right).\label{P2def}
\end{eqnarray}

A remarkable result is that in the superhorizon limit ($k\eta \to
0$) is that the phase of $\vchi_0(k,\eta)$ freezes and it reads,
\begin{equation}\label{freev02}
\vchi_0(k,\eta)=\frac{\ii}{\eta}\frac{1}{\sqrt{2k^3}}.
\end{equation}
Note that the Fourier modes of $\chi$ are frozen only in the case
of a massless field in a de Sitter background. As a result the
field $\chi$ behaves like a classic stochastic field with
fluctuations whose 2-point correlator is,
\begin{equation}\label{powerv0}
\mg \chi(\vk_1)\chi(\vk_2)\md= \frac{H^2}{2\,k_1^3}\ \delta_{\rm
Dirac}(\vk_1+\vk_2)
\end{equation}
And since $\chi$ is a free field, its superhorizon fluctuations
follow a Gaussian statistics.

\subsection{Computation of higher-order correlation functions}

Having determined the free field solutions, one can then express
perturbatively the $N$-point correlation functions of the interacting
field, $\chi$, in terms of those of the free scalar field. The equal
time correlators we are interested in are expectation values of
product of field operators for the current time vacuum state. Such
computations can be performed following general principles of quantum
field calculations~\cite{maldacena}. The simplest formulation it
is to apply the evolution operator $U(\eta_0,\eta)$ backward in time
to transform the interacting field vacuum into the free field vacuum
at an arbitrarily early time $\eta_0$ so that,
\begin{equation}\label{CumulantGeneral}
  \mg \vchi_{\vk_1}\ldots\vchi_{\vk_n}\md\equiv
  \mg 0\vert U^{-1}(\eta_0,\eta)\ \vchi_{\vk_1}\ldots\vchi_{\vk_n}\ U(\eta_0,\eta)\vert
  0\md
\end{equation}
where $\vert 0 \md$ is here the \emph{free field} vacuum~\footnote{It
is to be noted that these calculations do not correspond to those of
diffusion amplitudes of some interaction processes in a de Sitter
space. When one tries to do these latter calculations with a path
integral formulations, mathematical divergences are encountered as it
has been stressed in~\cite{tsamis,BU1}. With that respect de Sitter
space differs from Minkowski space-time. Our current point of view on
this difficulty is that diffusion amplitudes cannot be properly
defined in de Sitter space and that only field correlators as defined
here correspond to actual observable quantities that can be safely
computed.}. It is implicitly assumed in this expression that the
coupling of the field $\chi$ is switched on at time $\eta_0$. We will
see in the following that the choice of $\eta_0$ is not important as
long as it is much earlier than any other times intervening in the
problem.

The evolution operator $U$ can be written in terms of the interaction
Hamiltonian, $H_I$, as
\begin{equation}\label{Uoperator}
  U(\eta_0,\eta)=\exp{\left(-\ii \int_{\eta_0}^{\eta}\d\eta'\ H_I(\eta')\right)}
\end{equation}

If one is interested only in a single vertex interaction quantity, the
evolution operator can be expanded to linear order in $H_I$,
\begin{equation}\label{Uexpansion}
  U(\eta_0,\eta)=I_{\rm d}-\ii \int_{\eta_0}^{\eta}\d\eta'\ H_I(\eta')
\end{equation}
so that the connected part of the above ensemble average at a time
$\eta$ finally reads,
\begin{equation}\label{Cumulant}
\mg
  \vchi_{\vk_1}\ldots\vchi_{\vk_n}\md_c=-\ii\int_{\eta_0}^{\eta}\d\eta'\
  \mg 0\vert
  \left[\vchi_{\vk_1}\ldots\vchi_{\vk_n},H_I(\eta')\right]\vert 0\md
\end{equation}
where the brackets stand for the commutator.

The result is expressed in terms of the Green function
\begin{equation}\label{powerspectrum}
  G(k,\eta,\eta')=\frac{1}{2k}\left(1-\frac{\ii}{k\eta}\right)
  \left(1+\frac{\ii}{k\eta'}\right)\exp[\ii k(\eta'-\eta)]
\end{equation}
defined as $ \mg 0\vert v(\vk,\eta) v(\vk',\eta')\vert 0\md=
\delta_{\rm Dirac}(\vk+\vk')G(k,\eta,\eta')$.

 Before we proceed to explore explicit cases, let us note that as long
as calculations are restricted to tree order, the very same
calculation can be done assuming that $\chi$ is a classic stochastic
field whose stochastic properties are initially those of the quantum
free field. Not surprisingly one finds the same formal expressions!

\section{Exact results}

Results can be given in a closed form for simple self-interacting
potentials. We give in this section explicit results for cubic and
quartic potentials. Because of its renormalization properties, quartic
potential is a natural choice to consider. We will however see that
the case of a cubic potential can be relevant when finite volume
effects are taken into account (see Ref.~\cite{BU3} for more details
on finite volume effects).

In the following we thus assume that $H_I$ is of the form
\begin{equation}\label{HIexpression}
H_I(\chi)=\int\d^3\vx \sqrt{-g}\,{\lambda\over p!}\chi^p,
\end{equation}
and we are then interested in the computation of the leading order
part of the connected part of the ensemble average of products of
$p$ Fourier modes of the fields.

In case of a quartic coupling, we have,
\begin{equation}\label{fourthcumgeneral}
\mg \vchi_{\vk_1}\ldots\vchi_{\vk_4}\md_c = -\ii\lambda\delta_{\rm
Dirac}(\vk_1+\ldots+\vk_4) \int^\eta\d\eta'
 \left[G(k_1,\eta,\eta')\ldots
 G(k_4,\eta,\eta')-G^*(k_1,\eta,\eta')\ldots
 G^*(k_4,\eta,\eta')\right]
\end{equation}
This expression is unfortunately quite cumbersome to compute. It
depends, for symmetry reasons, on the norms of the four wave vectors
$k_1$,...,$k_4$ in the following combinations
\begin{eqnarray}
\pi_1&=&\sum_i k_i\\
\pi_2&=&\sum_{i < j} k_i\,k_j\\
\pi_3&=&\sum_{i < j < k} k_i\,k_j\,k_k\\
\pi_4&=&\sum_{i < j < k < l} k_i\,k_j\,k_k\,k_l
\end{eqnarray}
The expression (\ref{fourthcumgeneral}) can then be rewritten as
\begin{eqnarray}
\mg \vchi_{\vk_1}\ldots\vchi_{\vk_4}\md_c &=& \frac{-\lambda}
{24\,\pi_4^3\,\pi_1\,\eta^4}\,\delta_{\rm Dirac}\left(\sum_i\vk_i\right)
\Big(\nonumber\\
&&\hspace{-1cm}
      {\pi_1}^4 - 2\,{\pi_1}^2\,\pi_2 -
      \pi_1\,\pi_3 + 3\,\pi_4 + \left(-{\pi_1}^3\,\pi_3 +
      3\,\pi_1\,\pi_2\,\pi_3 - {\pi_1}^2\,\pi_4 -
      3\,\pi_2\,\pi_4\right)\eta^2 + 3\,{\pi_4}^2\,\eta^4\nonumber\\
&&\hspace{-1cm}-
      \pi_1\,\left( {\pi_1}^3 - 3\,\pi_1\,\pi_2 +
         3\,\pi_3 \right) \,\Big\{
       \left[ \left( 1 - \pi_2\,\eta^2 + \pi_4\,\eta^4 \right) \,\cos (\pi_1\,\eta) +
         \eta\,\left( \pi_1 - \pi_3\,\eta^2 \right) \,\sin (\pi_1\,\eta)
         \right]\,{\rm ci}(-\pi_1\,\eta)\nonumber\\
&&\hspace{-1cm}-
         \left[ \eta\,\left( \pi_1 - \pi_3\,\eta^2 \right)\,
          \cos (\pi_1\,\eta) + \left( -1 + \pi_2\,\eta^2 - \pi_4\,\eta^4 \right) \,
          \sin (\pi_1\,\eta) \right] \,{\rm si}(-\pi_1\,\eta) \Big\}\Big)\label{cum4expression}
\end{eqnarray}
where the CosIntegral (ci) and SinIntegral (si) functions are defined,
for $ \eta < \!0$, by
\begin{equation}
\int_{-\infty}^{\eta}\frac{\d\eta'}{\eta'}\exp[-\ii k \eta']
 \equiv
 {\rm ci}(-k\eta) + i\,{\rm si}(-k\eta).
\end{equation}

The superhorizon limit (i.e.  $k_i\eta\ll 1$ for all $i=1\ldots4$)
of Eq.~(\ref{cum4expression}) is
\begin{eqnarray}
\mg \vchi_{\vk_1}\ldots\vchi_{\vk_4}\md_c &=&
\frac{-\lambda\,\,\delta_{\rm
Dirac}\left(\sum_i\vk_i\right)}{24\,\pi_4^3\,\pi_1\,\eta^4}\Big\{{\pi_1}^4 -
2\,{\pi_1}^2\,\pi_2 - \pi_1\,\pi_3 + 3\,\pi_4
      -\pi_1\,\left( {\pi_1}^3\!-\!3\,\pi_1\,\pi_2\!+\!3\,\pi_3 \right) \,
       \left[\gamma+\log (-\pi_1\,\eta)\right]  \Big\}
\end{eqnarray}
where $\gamma$ is the Euler's constant ($\gamma\approx 0.577$).  This
implies that the 4-point correlator of the actual fields $\chi$ reads
\begin{equation}
\mg \chi_{\vk_1}\ldots\chi_{\vk_4}\md_c = -\ \frac{\lambda\ H^4
}{24}\frac{\delta_{\rm Dirac}(\sum \vk_i)}{ \prod k_i^3}\
\left[-\sum k_i^3\left(\gamma+\zeta(\{k_i\})+\log\left[-\eta\sum
k_i\right]\right) \right]\label{fourthcumlimit}
\end{equation}
In this expression, terms of the order of $k_i\eta$ have been
neglected. This result illustrates the transition to the
stochastic limit. The function $\zeta$ is an homogeneous function
of the Fourier wave-numbers
\begin{equation}\label{zeta}
  \zeta(\{k_i\})=\frac{-{\pi_1}^4 +
2\,{\pi_1}^2\,\pi_2 + \pi_1\,\pi_3 - 3\,\pi_4}{\pi_1\,\left(
{\pi_1}^3 - 3\,\pi_1\,\pi_2 + 3\,\pi_3 \right)}.
\end{equation}

\begin{figure}
\centerline{ \psfig{figure=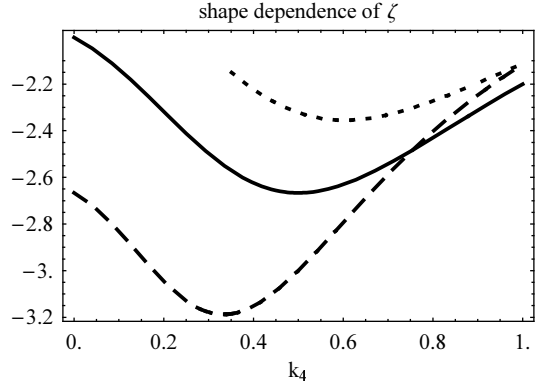,width=7cm}} \caption{
Dependence of the function $\zeta$ as a function of $k_4$ for
fixed values of $k_1$, $k_2$ and $k_3$. The solid line corresponds
to configuration $k_1$=0, $k_2$=$\frac{1}{2}$,
$k_3$=$\frac{1}{2}$, $k_4$; the dashed line to
$k_1$=$\frac{1}{3}$, $k_2$=$\frac{1}{3}$, $k_3$=$\frac{1}{3}$,
$k_4$ and the dotted line to $k_1$=$\frac{1}{6}$,
$k_2$=$\frac{1}{3}$, $k_3$=$\frac{1}{6}$, $k_4$. }
  \label{zetatot}
\end{figure}

The dependence of the function $\zeta$ on the wavelength ratios is
illustrated on Fig. \ref{zetatot}. It is found to be relatively
weak. We did not fully explore its properties but one can explicitly
show for instance that if one of the wave vectors vanishes, $\zeta$ is
smaller than $-2$ and reaches its minimum ,$-{8}/{3}$, for a symmetric
configuration of the three remaining wave vectors. From the plot on
Fig.~\ref{zetatot} it appears that the minimum value of $\zeta$ is
$-3.2$ and is reached for a ''square" configuration, that is when the
four wavelengths are of equal length. The maximum value, $-2$, is
reached when two of the wavelengths vanish. The overall variations of
$\zeta$ with the wavelength ratios are therefore rather mild. It is
then legitimate to describe the four-point correlation function as the
sum of products of two-point power spectra, e.g.,
\begin{eqnarray}
  \mg \chi_{\vk_1}\ldots\chi_{\vk_4}\md_c &=&
      \delta_{\rm Dirac}\left(\sum \vk_i\right)\,P_4(\{k_i\})\\
  P_4(\{k_i\})&=&\nu_3(\{k_i\})\sum_{i} \prod_{j \ne i}
  \frac{H^2}{2 k_j^3}.\label{p4sh}
\end{eqnarray}
The vertex value, $\nu_3$ is given by
\begin{equation}\label{vertexvalue}
  \nu_3(\{k_i\})=\frac{\lambda}{3 H^2}\left[\gamma+\zeta(k_i)
  +\log\left(-\eta\sum k_i\right)\right]
\end{equation}
that carries a weak geometrical dependence with the wave vectors
geometry through $\zeta$.

When the term $\log\left(-\eta\sum k_i\right)$ is large (and
negative), that is when the number of $e$-folds, $N_e$, between the
time of horizon crossing for the modes we are interested in and the
end of inflation is large, the vertex value is simply given
$$
\nu_3(\{k_i\})=-{\lambda\,N_e}/{(3\,H^2)}
$$
which corresponds exactly to the value that was obtained in
Ref.~\cite{BU1} from the classical evolution of the stochastic field
on superhorizon scales (see comments in last section).

\begin{figure}
\centerline{ \psfig{figure=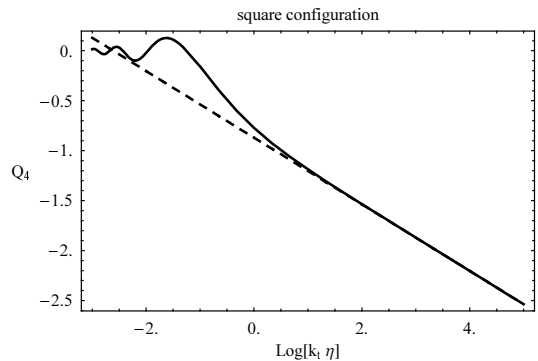,width=7cm}}
\caption{Behaviour of the function $Q_4$ as of function of time.  The
transition to the superhorizon behavior (dashed line) is shown. The
function $Q_4$ is shown here for a ''square" configuration
($k_1=k_2=k_3=k_4$) as a function of $k_t\eta=\sum k_i\eta$.}
\label{Q4square}
\end{figure}

\begin{figure}
  \centerline{ \psfig{figure=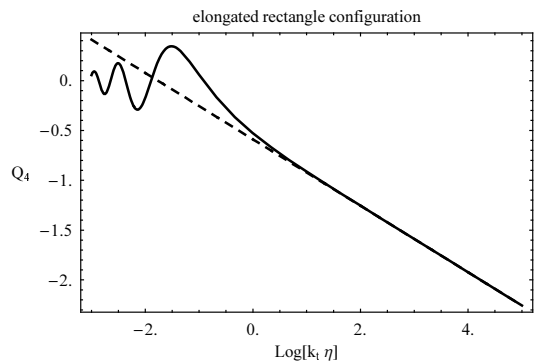,width=7cm}}
\caption{Same as Fig.~\ref{Q4square} for a ''rectangular"
configuration ($k_1=k_2=4,\, k_3=4 k_4$).}
  \label{Q4rectangle}
\end{figure}

\begin{figure}
  \centerline{ \psfig{figure=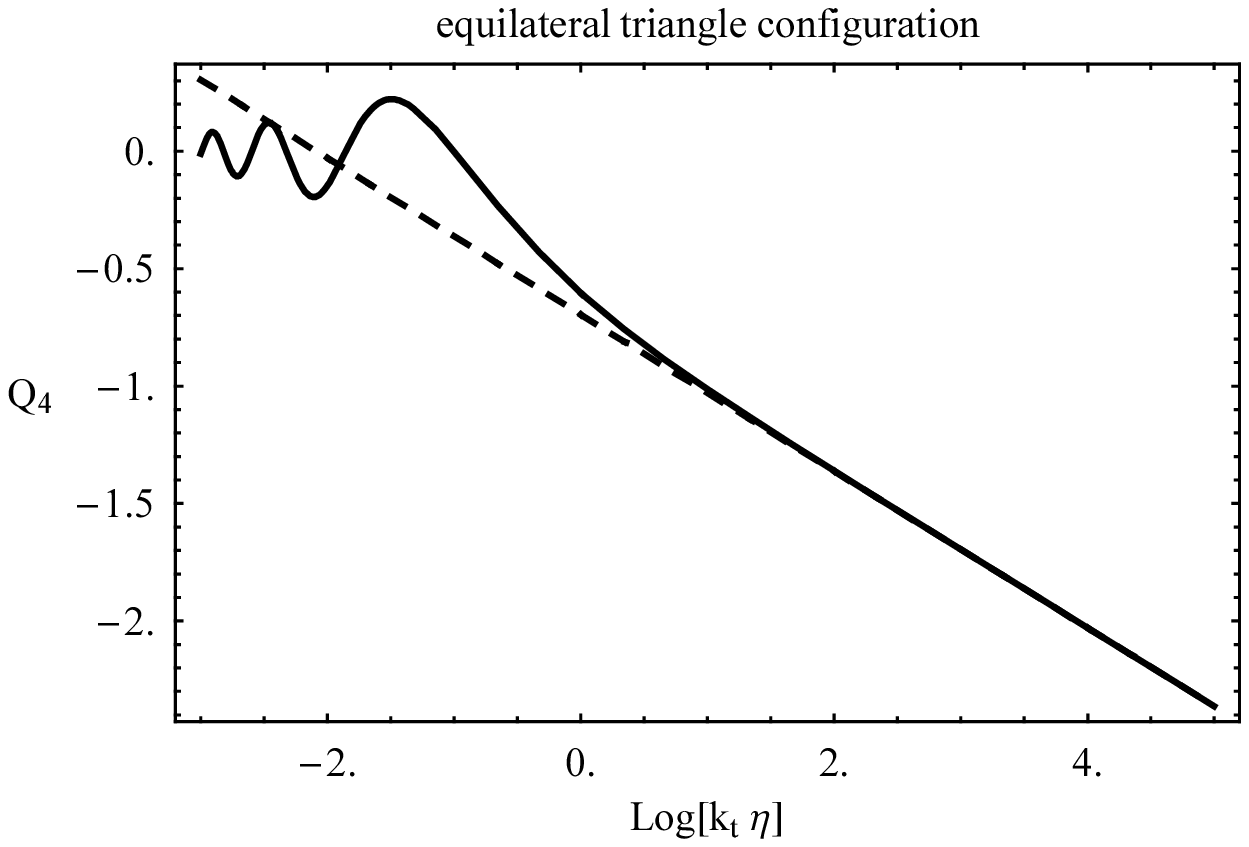,width=7cm}}
\caption{Same as Fig.~\ref{Q4square} for a ''triangular"
configuration ($k1=0, k_2=k_3=k_4$).}
  \label{Q4triangle}
\end{figure}

The result obtained above is however more complete since it contains
next to leading order terms. It allows, for instance, to estimate the
validity regime of the superhorizon result. It shows in particular
that the mode coupling induced on subhorizon scales are negligible as
soon as the number of $e$-folds after horizon crossing exceeds a few
units. This is made clear when one considers the reduced four-point
correlation function, $Q_4$, defined as,
\begin{equation}\label{q4}
  Q_4(\{k_i\})=\frac{P_4(k_1,k_2,k_3,k_4)}{P_2(k_1)P_2(k_2)P_2(k_3)+{\rm
  sym.}}
\end{equation}
where $P_2$ is defined in Eq.~(\ref{P2def}). The behaviour of $Q_4$ as
a function of time and for different configurations of the wavevectors
is depicted on Figs.~(\ref{Q4square}-\ref{Q4triangle}) and one can
convince hismself that it converges rapidly toward the superhorizon
result as soon as $-\log\sum k_i\eta \sim {\cal O}(1)$.

Similar results can be obtained for the 3-point function in the case
of a cubic potential, $V(\chi)=\lambda\,\chi^3/3!$. In this case the
quantity to compute is
\begin{equation}\label{thirdcumgeneral}
\mg \vchi_{\vk_1}\ldots\vchi_{\vk_3}\md = -\ii\lambda\delta_{\rm
Dirac}(\vk_1+\ldots+\vk_3) \int^\eta\frac{-\d\eta'}{H\eta'}
 \left[G(k_1,\eta,\eta')\ldots
 G(k_3,\eta,\eta')-G^*(k_1,\eta,\eta')\ldots
 G^*(k_3,\eta,\eta')\right].
\end{equation}
which gives,
\begin{eqnarray}
\mg \vchi_{\vk_1}\ldots\vchi_{\vk_3}\md &=& \frac{\lambda}
{12\,\pi_3^3\,\eta^3\,H}\,\delta_{\rm Dirac}(\vk_1+\ldots+\vk_3) \Big(\nonumber\\
&&\hspace{-1cm}
      {\pi_1}^3 - 2\,{\pi_1}\,\pi_2 -
      \pi_3 + \left(-{\pi_1}^2\,\pi_3 +
      3\,\pi_2\,\pi_3\right)\eta^2 \nonumber\\
&&\hspace{-1cm}-
      \left( {\pi_1}^3 - 3\,\pi_1\,\pi_2 +
         3\,\pi_3 \right) \Big\{ \,
       \left[ \left( 1 - \pi_2\,\eta^2 \right) \,\cos (\pi_1\,\eta) +
         \eta\,\left( \pi_1 - \pi_3\,\eta^2 \right) \,\sin (\pi_1\,\eta)
         \right]\,{\rm ci}(-\pi_1\,\eta)\nonumber\\
&&\hspace{-1cm}-
         \left[ \eta\,\left( \pi_1 - \pi_3\,\eta^2 \right)\,
          \cos (\pi_1\,\eta) + \left( -1 + \pi_2\,\eta^2 \right) \,
          \sin (\pi_1\,\eta) \right] \,{\rm si}(-\pi_1\,\eta) \Big\}\Big)\label{cum3expression}
\end{eqnarray}
To a factor $2\eta$, one recovers basically the same expression as for
the fourth cumulant (\ref{cum4expression}) when $\pi_4$ is set to
zero. In the superhorizon limit, the expression of this cumulant reads
\begin{equation}
\mg \chi_{\vk_1}\ldots\chi_{\vk_3}\md = -\ \frac{\lambda\ H^2
}{12}\frac{\delta_{\rm Dirac}\left(\sum \vk_i\right)}{ \prod k_i^3}\
\left[-\sum k_i^3\left(\gamma+\zeta_3(\{k_i\})+\log\left[-\eta\sum
k_i\right]\right) \right]\label{thirdcumlimit}
\end{equation}
where $\zeta_3$ is simply given by
\begin{equation}\label{zeta3exp}
  \zeta_3(\{k_i\})=\zeta(k_1,k_2,k_3,k_4\to 0).
\end{equation}
To express it in another way we still have
\begin{eqnarray}
  \mg \chi_{\vk_1}\ldots\chi_{\vk_3}\md &=& \delta_{\rm Dirac}\left(\sum
  \vk_i\right)\,P_3(\{k_i\})\\
  P_3(\{k_i\})&=&\nu_3(\{k_i\})\sum_{i} \prod_{j \ne i}
  \frac{H^2}{2 k_j^3}\label{p3sh}
\end{eqnarray}
with a vertex value $\nu_3$ to be taken in the appropriate limit,
$k_4\to 0$. The shape dependence of the vertex, or the time
dependance of this cumulant thus reproduces that of the fourth
cumulant as shown on Figs. \ref{zetatot} (solid line) and
\ref{Q4triangle}.

\section{Comments}

We have obtained some closed forms for the 3- and 4-point correlation
functions of a test scalar field with a self-interacting potential of
order respectively 3 and 4 in a de Sitter background.

It is interesting to note, as it was already in the
literature~\cite{BU1}, that the superhorizon behaviour of the field
can be obtained from a simplified Klein-Gordon equation for the field
evolution on superhorizon scales
\begin{equation}\label{chievol}
  \ddot{\chi}+3H\dot\chi=-\frac{\partial V}{\partial \chi}(\chi)
\end{equation}
solved perurbatively at first order in $\lambda$. In this equation
$\chi$ has to be understood as the filtered value of $\chi$ at a fixed
scale that leaves the horizon at a given time $t_0$. The previous
equation is then valid for $t>t_0$ only where the field can be
described by a classical stochastic field. Moreover in writing this
equation one also makes the assumption that its r.h.s. can be computed
from the filtered value of the field (which is not necessarily
identical to what would have been obtained from a filtering of the
source term). The equation (\ref{chievol}) can be solved
perturbatively in $\lambda$. At zeroth order $\chi$ is constant and, at
first order, it reads
\begin{equation}\label{chi1}
  \chi^{(1)}=\chi_0^{(1)}-\frac{\partial V}{\partial
  \chi}\left(\chi^{(0)}\right)\frac{t-t_0}{3\,H},
\end{equation}
if $\chi_0^{(1)}$ is the leading order value of the field at horizon
crossing. Note that $t-t_0$ can be rewritten as $N_e/H$ where $N_e$ is
the number of $e$-folds since horizon crossing. It implies that if the
number of $e$-folds is large enough the term $\chi_0^{(1)}$ should
become negligible. As a consequence the leading order expression of
the first nontrivial high order cumulant of the one-point PDF of
$\chi$ takes either the form $3\,\nu_3\mg\chi^2\md^2$ or
$4\,\nu_3\mg\chi^2\md^3$ for respectively a cubic or a quartic
potential. The value of the coefficient $\nu_3$ that enters these
expressions is precisely the one obtained in Eq.~(\ref{vertexvalue})
in the superhorizon limit, that is
$\nu_3=-{\lambda\,N_e}/{(3\,H^2)}$. And indeed if one had to compute
such cumulants from either expression (\ref{p4sh}) or (\ref{p3sh}) the
integration over the wave vectors would have led to the very same
expressions in the superhorizon limit. That shows that the late time
behavior of the cumulants we found comes in fact from a simpler
dynamical evolution. It actually demonstrates that the computations in
the classical approach sketched here can be put on a firm ground. It
also allows investigations of more subtle effects, such as the finite
volume effects where more than one filtering scales have to be taken
into account, that cannot be properly addressed in a classical
approach.

Regarding finite volume effect it is interesting to investigate how
the results for the cubic and the quartic potentials could be related
together. We refer here to our companion paper~\cite{BU3} where we
give a comprehensive presentation of finite volume effects on
observable quantities. Suffice is to say that even if one assumes
$\chi$ has a self interacting quartic potential, a non-zero third
order correlator might be observable. The reason is that one cannot
have access to genuine ensemble averages but to constrained ensemble
averages such as $\mg \chi_{\vk_1}\dots\chi_{\vk_3}\md_{\chib}$ which
is the expectation value of $\chi_{\vk_1}\dots\chi_{\vk_3}$ for a
given value of $\chib$, average value of $\chi$ over the largest scale
available in the survey in which the correlators are computed.

To leading order in $\chib$ with respect to its variance,
$\sigma_{\chib}$, this constrained average value reads
\begin{equation}\label{cv3}
  \mg \chi_{\vk_1}\dots\chi_{\vk_3}\md_{\chib}=
  \mg \chi_{vk_1}\dots\chi_{\vk_3}\,\chib\md_c\,\frac{\chib}{\sigma^2_{\chib}}.
\end{equation}
If, for $i=1\ldots3$, the wavelengths $1/k_i$ are much smaller than
the survey size then the expression of
$\mg\chi_{\vk_1}\dots\chi_{\vk_3}\,\chib\md_c\,{\chib}/{\sigma^2_{\chib}}$
is that of $\mg\chi_{\vk_1}\dots\chi_{\vk_3}\md$ when the
self-interacting field $\chi$ evolves in the potential
${\lambda}\chib\,\chi^3/3!$, which is the cubic term in $\chi$ in the
expansion of $\lambda\,(\chi+\chib)^4/4!$ . This shows that the two
cases described in this paper are consistent with one another and that
the functional relation between the two had to be expected.

To conclude we have established in this paper a number of generic
results that put the tree-order computation of higher order
correlations in a de Sitter background on a secure ground. We
leave for further studies the examination of more realistic cases
in which the background expansion is more complex.

\end{document}